\documentclass[showpacs,prl,aps,twocolumn]{revtex4}
\usepackage{amsmath,amsfonts,bm,graphicx,hyperref}

\newcommand{\llangle}{\langle\!\langle}
\newcommand{\rrangle}{\rangle\!\rangle}

\begin{document}
\title{Accuracy of the quantum capacitor as a single-electron source}
\author{Mathias Albert}
\author{Christian Flindt}
\author{Markus B\"uttiker}
\affiliation{D\'epartement de Physique Th\'eorique, Universit\'e de Gen\`eve, CH-1211 Gen\`eve, Switzerland}
\date{\today}

\begin{abstract}
A periodically driven quantum capacitor may function as an on-demand single electron source as it has recently been demonstrated experimentally. However, the accuracy at which single electrons are emitted is not yet understood. Here we consider a conceptually simple model of a quantum capacitor and find analytically the noise spectrum as well as the counting statistics of emitted electrons. We find that the failure rate of the capacitor can be arbitrarily small when operated under favorable conditions. Our theoretical predictions may be tested in future experiments.
\end{abstract}



\maketitle

\textit{Introduction}.--- Controllable single electron sources are at the forefront of current research on nano-scale electronics. Systems that generate quantized electrical currents, for example quantum capacitors \cite{Fev07,Mos08,Kee08} and quantum pumps \cite{Blu07}, are of great interest due to their potential applications in metrology \cite{Fle99} and quantum information processing \cite{Bee03} as well as in basic research on single- and few-electron physics in mesoscopic structures. The quantum capacitor constitutes one archetype of a single electron emitter. The capacitor consists of a nano-scale cavity that exchanges particles with a reservoir through a narrow constriction, a so-called quantum point contact. When the capacitor is subject to periodic voltage modulations, single electron emission and absorption occur at giga-hertz frequencies as it has recently been demonstrated experimentally \cite{Fev07}. It has also been verified \cite{Gab06} that the relaxation resistance of the capacitor is quantized in units of $h/2e^2$ independently of microscopic details, in agreement with theoretical predictions \cite{But93}. The experiments indicate that the quantum capacitor may function as an on-demand electron source that ideally can be controlled down to the level of single electrons.

Despite the experimental and theoretical advances, the accuracy at which the quantum capacitor emits electrons is still not well understood. This is an important question for potential applications and it may ultimately be the criterion that determines if the quantum capacitor becomes an integrated part of future nano-scale electronics operating at giga-hertz frequencies. In this work we analyze the accuracy of the quantum capacitor as a single electron source. To this end, it is necessary not only to study the mean current of electrons emitted by the capacitor but also the fluctuations of the current around the mean. We describe the capacitor using a simple model which has been shown to reproduce noise measurements in numerical simulations of the recent experiment reported in Ref.\ \cite{Mah10}. We derive an analytic expression for the noise spectrum that fully accounts for the measurements. Furthermore, we characterize fluctuations in the current of emitted electrons by evaluating the counting statistics and find that the failure rate of the device under optimal operating conditions can be vanishingly small.

\begin{figure}
  \begin{center}
    \includegraphics[width=0.95\linewidth]{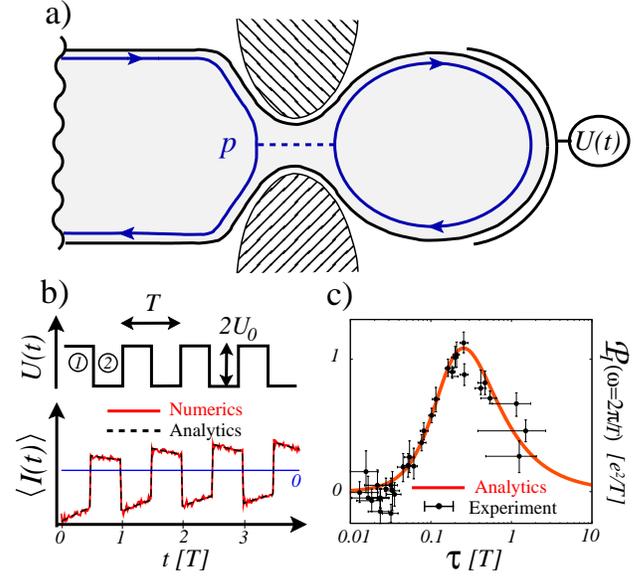}
    \caption{Quantum capacitor. {\bf a)} An edge state is connected to the capacitor via a quantum point contact. The periodic voltage $U(t)$ applied to the capacitor causes emission and absorption of single electrons to and from the edge state. {\bf b)} Periodic voltage $U(t)$ and resulting current $\langle I(t)\rangle$ as functions of time. Numerical and analytical results are shown. {\bf c)} Analytic result for the noise spectrum $\mathcal P_I(\omega=2\pi/T)$ as function of the correlation time $\tau$ (see text). Experimental results have been adapted from Ref.\ \cite{Mah10}.}
    \label{fig:system}
  \end{center}
\end{figure}

\textit{Quantum capacitor}.--- The system is shown schematically in Fig.~\ref{fig:system}a. An edge state is connected to the capacitor via a quantum point contact whose transmission probability $p$ can be controlled with external gates. With the quantum point contact pinched off, the capacitor has a discrete energy spectrum with typical level spacing $\Delta$, which is much larger than temperature, and $\tau_o= h/\Delta$ is the time it takes an electron to travel a full round along the edge of the capacitor. As the quantum point contact is opened, electronic states of the capacitor with energies below the Fermi level of the external reservoir are filled. We now consider the situation where a step-like gate voltage $U(t)$ periodically shifts the highest occupied level above and below the Fermi level of the reservoir. The period $T$ is much longer than $\tau_o$ and the amplitude $2U_0$ is on the order of the level spacing $\Delta/e$. This causes periodic emission of coherent single electron wave packets from the capacitor to the outgoing edge state, followed by refilling from the incoming edge state as it was recently demonstrated experimentally \cite{Fev07}. In this work, we study the accuracy of single electron emission as the ratio of the period $T$ and the escape, or correlation, time $\tau$ (defined below) is varied or, equivalently, as function of the dimensionless parameter $\varepsilon=e^{-T/2\tau}$.

\textit{Model}.--- The charge dynamics of the capacitor follows a simple model that recently was shown to reproduce measured data in numerical simulations of the experiment reported in Ref.\ \cite{Mah10}. The absorption phase of duration $T/2$ (denoted by \textcircled{\footnotesize{1}} in Fig.~\ref{fig:system}b) is discretized in time steps of length $\tau_o$ during each of which a single electron can enter the capacitor with probability $p$. The emission phase (denoted by \textcircled{\footnotesize{2}}) is similarly discretized in time, and in each time step the probability of emitting an electron is $p$. With the amplitude of the periodic driving being on the order of the level spacing, higher-lying states can safely be neglected and only a single (additional) electron can occupy the capacitor. This semi-classical picture can be formulated as a master equation in discrete time for the probability of the capacitor to be occupied by an electron. Setting the electron charge $e=1$ in the following, this probability is equal to the average (additional) charge of the capacitor $\langle Q\rangle$, where $Q=0,1$. The master equation determines the evolution of the average charge after one time step and reads \cite{Mah10}
\begin{equation}\label{master_eq}
  \langle Q(t_{k+1,l})\rangle= \left\{\begin{array}{ll}
     p[1\!-\!\langle Q(t_{k,l})\rangle]\!+\!\langle Q(t_{k,l})\rangle & \textcircled{\footnotesize{1}} \\
      \\
     (1-p)\langle Q(t_{k,l})\rangle & \textcircled{\footnotesize{2}}
    \end{array}\right.
\end{equation}
where we have used that $1-\langle Q\rangle$ is the probability for the capacitor to be empty and $t=t_{k,l}$ denotes time at the $k$'th time step during the $l$'th period. The absorption (emission) phase $\textcircled{\footnotesize{1}}$ ($\textcircled{\footnotesize{2}}$) corresponds to $k=1,2\ldots,K$ ($K+1,K+2,\ldots,2K$). Although, we do not derive the master equation here, some insights into its origin can be found by considering the current as it was calculated in Ref.~\cite{Mos08} using scattering matrices. The current was shown to consist of
one step-like term with step length $\tau_o$ and one oscillatory part with period $\tau_o$. The oscillatory part is due to quantum interference and vanishes with increasing temperatures. At arbitrary temperature, only the first term remains after integration of the current over the time step $\tau_o$, which leads to the master equation above.  As in the experiment \cite{Mah10} (with the measurement frequency equal to the driving frequency $2\pi/\omega=T= 60\,\tau_o$), we consider time scales that are much longer than $\tau_o$ for which the master equation provides a quasi-continuous description.

\textit{Average charge and current}.---
It is straightforward to solve the master equation (\ref{master_eq}) for the average charge $\langle Q\rangle$ and thus the net current running out of the capacitor $\langle I(t)\rangle\equiv -\langle \dot{Q}(t)\rangle\simeq[\langle Q(t)\rangle-\langle Q(t+\tau_o)\rangle]/\tau_o$. The average charge can be cast in a form similar to that of an $RC$ circuit reading
\begin{equation}\label{qt2}
  \langle Q(t_{k,l})\rangle= \left\{\begin{array}{ll}
      \displaystyle 1-\beta_l\, e^{-(t_{k,l}-lT)/\tau} & \textcircled{\footnotesize{1}}  \\
       \\
      \displaystyle \alpha_l\, e^{-(t_{k,l}-[l+\frac{1}{2}]T)/\tau} & \textcircled{\footnotesize{2}}
    \end{array}\right.
\end{equation}
where we have defined $\alpha_l=1/(1+\varepsilon)+\theta \varepsilon^{2l}$ and $\beta_l=1/(1+\varepsilon)-\theta \varepsilon^{2l-1}$ with $\theta$ depending on the initial conditions at the time when the periodic driving is turned on. The correlation time $\tau\equiv\tau_o/\ln[1/(1-p)]$ determines the time scale over which the system loses memory about the initial conditions encoded in $\theta$ and $\langle Q\rangle$ becomes periodic in time. Figure \ref{fig:system}b illustrates the excellent agreement between the current $\langle I(t)\rangle$ obtained from the analytic expression (\ref{qt2}) and our numerical simulations based on Eq.~(\ref{master_eq}), including the initial transient behavior. The mean charge emitted during the emission phase is $\tanh\left(T/4\tau\right)$ [see also Eqs.\ (\ref{pnotw}) and (\ref{eq:cumulants})].

\textit{Noise spectrum}.--- In the experiment, the Fourier transform of the time-averaged correlation function $\overline{\langle \delta I(t)\delta I(t+t_0)\rangle}^{\,t}$ with $\delta I(t)=I(t)-\langle I(t)\rangle$  was measured \cite{Mah10}. The noise spectrum is then $\mathcal P_I(\omega)=\int_{-\infty}^{+\infty}dt_0\, \overline{\langle \delta I(t)\delta I(t+t_0)\rangle}^{\,t} e^{i\omega t_0}$ or $\mathcal P_I(\omega)\simeq\omega^2\mathcal P_Q(\omega)$ in terms of the corresponding charge correlation function $\mathcal P_Q(\omega)$. We evaluate the charge correlation function by noting that $\langle Q(t)Q(t+t_0)\rangle$ is the joint probability for the capacitor to be charged with one electron at time $t$ and at time $t+t_0$. Using conditional probabilities we write $\langle Q(t)Q(t+t_0)\rangle=\langle Q(t)\rangle \langle \widetilde Q(t+t_0)\rangle$, where $\langle \widetilde Q(t+t_0)\rangle$ is the probability that the capacitor is charged with one electron at time $t+t_0$ given that it is charged at time $t$. For $t_0>0$, $\langle \widetilde Q(t+t_0)\rangle$ can be found by propagating forward in time the condition $\langle \widetilde Q(t)\rangle=1$ using the master equation (\ref{master_eq}). Similar reasoning applies to the case  $t_0<0$.  Integrating over $t$, the time-averaged charge correlation function becomes $\overline{\langle\delta Q(t)\delta Q(t+t_0)\rangle}^{\,t}=\,\frac{\tau}{T}\,e^{-|t_0|/\tau}\tanh\left(\frac{T}{4\tau}\right)$, and
finally, we obtain the noise spectrum
\begin{equation}\label{pnotw}
  \mathcal P_I(\omega)=\frac{2}{T}\tanh\left(\frac{T}{4\tau}\right)\,\frac{\omega^2\tau^2}{1+\omega^2\tau^2}.
\end{equation}

Figure \ref{fig:system}c shows our analytic expression for the noise spectrum together with experimental results adapted from Ref.\ \cite{Mah10}. The analytic result captures the experiment over the full range of correlation times $\tau$ and interpolates between the two limiting cases discussed in Ref.\ \cite{Mah10}. In the shot noise regime $\tau\gg T$ ($\varepsilon \simeq 1$), the probability of emitting and reabsorbing an electron during a period is very small, and electron emission becomes rare. In this regime, we find $\mathcal P_I(\omega)\rightarrow 1/2\tau$ in agreement with Ref.\ \cite{Mah10}. In the phase noise regime $\tau\ll T$ ($\varepsilon\simeq 0$), the probability of emitting and absorbing an electron during each period is close to one, and the main source of finite-frequency fluctuations is the random times of emission and absorption within a period. In this regime we find $\mathcal P_I(\omega)\rightarrow \frac{2}{T}\frac{\omega^2\tau^2}{1+\omega^2\tau^2}$ as suggested in Ref.\ \cite{Mah10}. The zero-frequency limit $\mathcal P_I(0)=0$ reflects that charge does not accumulate on the capacitor over time once $\langle Q\rangle$ has become periodic in time.

\textit{Counting statistics}.--- While the noise spectrum (\ref{pnotw}) is related to the net current running out of the capacitor (including both absorption and emission of electrons), it is relevant to characterize the emission process alone in order to quantify the accuracy of the capacitor as a single electron source. To this end, we consider the counting statistics of emitted electrons and hence introduce the probabilities $P_j(n,l)$ that the capacitor is occupied by $j=0,1$ (additional) electrons, while $n$ electrons have been emitted after $l$ periods. The probability distribution for the number of emitted electrons is $P(n,l)=P_0(n,l)+P_1(n,l)$ and the corresponding cumulants $\llangle n^m\rrangle$ of the distribution are defined through the cumulant generating function (CGF) $\mathcal{S}(\chi,l)\equiv  \ln\left[\sum_n P(n,l)e^{in\chi}\right]$ as $\llangle n^m\rrangle=\partial^m_{(i\chi)}\mathcal{S}(\chi,l)|_{\chi\rightarrow 0}$. The CGF can be written $\mathcal{S}(\chi,l)=\ln\left[\mathbf{1}\cdot \mathbf{P}(\chi,l)\right]$, where $\mathbf{1}=[1,1]^T$ and $\mathbf{P}(\chi,l)=[P_1(\chi,l),P_0(\chi,l)]^T$ with  $P_j(\chi,l)=\sum_n P_j(n,l)e^{in\chi}$, $j=0,1$. The evolution of the probability vector after one period of the driving is obtained from the master equation (\ref{master_eq}) and reads $\mathbf{P}(\chi,l+1)=\mathbf{A}(\chi)\, \mathbf{P}(\chi,l)$ with $\mathbf{A}(\chi)=\mathbf{L}_1^{\frac{T}{2\tau_o}} \mathbf{L}_2^{\frac{T}{2\tau_o}}(\chi)$, where $\mathbf{L}_1=\left(\begin{array}{cc}1 & p\\ 0 & 1-p\end{array}\right)$ and $\mathbf{L}_2(\chi)=\left(\begin{array}{cc} 1-p & 0\\ p\,e^{i\chi} & 1 \end{array}\right)$, and we have introduced the counting field $\chi$ in the off-diagonal element of $\mathbf{L}_2$ corresponding to electron emission \cite{Bar03}. The CGF after $l$ periods is then $\mathcal{S}(\chi,l)=\ln\left[\mathbf{1}\cdot\mathbf{A}^l(\chi)\mathbf{P}_{\mathrm{in}}\right]$ with $\mathbf{P}_{\mathrm{in}}$ being the $\chi$-independent initial condition as counting begins. For a large number of periods, $\mathbf{A}^l(\chi)$ is dominated by the largest eigenvalue of $\mathbf{A}(\chi)$
\begin{equation}
  \lambda (\chi)=\varepsilon+ e^{i\chi}\,\frac{1-\varepsilon}{2}\left[1-\varepsilon+\sqrt{(1-\varepsilon)^2+4\varepsilon e^{-i\chi}}\right]
\end{equation}
such that $\mathcal{S}(\chi,l)\rightarrow l \ln[\lambda(\chi)]$ \cite{Pis04}. The expression for the CGF is a powerful result that allows us to fully characterize fluctuations in the current of emitted electrons.

\textit{Cumulants}.--- The cumulants of the current are defined as the constant ratio $\llangle I^m\rrangle=\llangle n^m\rrangle/l$ after a large number of periods. The first three cumulants are
\begin{equation}
\label{eq:cumulants}
\begin{split}
  \llangle I\rrangle &= \frac{1-\varepsilon}{1+\varepsilon}=\tanh\left(T/4\tau\right),\\
  \llangle I^2 \rrangle &= \frac{2\varepsilon}{(1+\varepsilon)^2} \llangle I\rrangle,\\
  \llangle I^3 \rrangle &= \frac{2\varepsilon(4\varepsilon-\varepsilon^2-1)}{(1+\varepsilon)^4}\llangle I\rrangle.
  \end{split}
\end{equation}
Higher-order cumulants can be approximated by noting that the CGF has square-root branch points at $i\chi_\pm=\ln[(1-\varepsilon)^2/4\varepsilon]\pm i\pi$, close to which the CGF behaves as $\mathcal{S}(\chi,l)\simeq 2l\sqrt{i\chi_\pm-i\chi}$. Following Ref.\ \cite{Fli09} we find for large orders $\llangle I^m \rrangle \simeq \frac{4B_{m,-1/2}}{|i\chi_+|^{m-1/2}}\cos{\left[(m-1/2)\arg(i\chi_+)\right]}$ with $B_{m,\mu}=\mu(\mu+1)\ldots(\mu+m-1)$. The asymptotic expression gives excellent agreement with exact results for the high-order cumulants as we have checked.

\begin{figure}
  \begin{center}
    \includegraphics[width=0.87\linewidth]{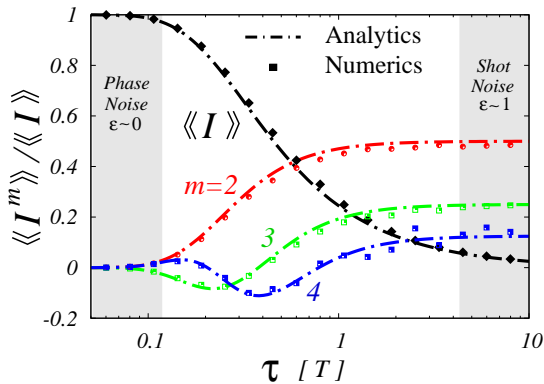}
    \caption{\label{figcumulant} Mean current and cumulants. Mean current $\llangle I\rrangle$ together with the normalized cumulants of the current $\llangle I^m\rrangle/\llangle I\rrangle$, $m=2,3,4$, as functions of the correlation time $\tau$ or, equivalently, $\varepsilon=e^{-T/2\tau}$. Symbols indicate results obtained from numerical simulations (obtained from 50.000 realizations), while dashed lines show our analytic expressions.}
  \end{center}
\end{figure}

Figure \ref{figcumulant} illustrates the good agreement between numerical simulations of the first four cumulants and our analytical results. The first cumulant (the mean current) is equal to the average charge emitted during one period which varies from 0 in the shot noise regime ($\varepsilon\simeq 1$) to 1 in the phase noise regime ($\varepsilon\simeq 0$). In the shot noise regime the first few cumulants can be written as $\llangle I^m\rrangle\simeq(1/2)^{m-1}\llangle I\rrangle$, indicating that the transport statistics is similar to that of a Poisson process with an effective charge of $1/2$ -- this is characteristic for a Poisson process in which only every second event results in emission. In the phase noise regime, the mean current is close to 1 and the first few cumulants are close to zero since electrons are emitted in an orderly manner due to the periodic driving. Low-frequency fluctuations in the stream of emitted electrons arise only in the rare cases when the capacitor is not charged in the absorption phase or when it fails to emit in the emission phase. We predict the occurrence of such failures in the phase noise regime by expanding the CGF to lowest order in $\varepsilon$ as $\mathcal{S}(\chi,l)\simeq l[i\chi+2\varepsilon(e^{-i\chi}-1)]$. The first term corresponds to a deterministic process in which one electron is emitted in each cycle. The second term is the sum of two independent Poisson processes describing ``cycle missing'' events occurring with rate $\varepsilon$ per period, either because the capacitor is not charged or because it fails to emit. Of course, if the capacitor is not charged it also fails to emit, however, such correlations do not enter to lowest order in $\varepsilon$. In the cross-over from the shot noise to the phase noise regime, the third cumulant changes from positive to negative before it vanishes in the phase noise regime. The negative third cumulant signals a left-skewed distribution caused by the mean current $\llangle I\rrangle$ being close to the upper limit of 1, which cuts off the distribution to the right. The fourth cumulant goes through a region with negative values followed by a region with positive values before vanishing in the phase noise regime. The negative (positive) fourth cumulant indicates a sub-gaussian (super-gaussian) distribution with light (heavy) tails. Higher-order cumulants become increasingly oscillating functions of $\varepsilon$ as recently predicted \cite{Fli09}.

\begin{figure}
  \begin{center}
    \includegraphics[width=0.86\linewidth]{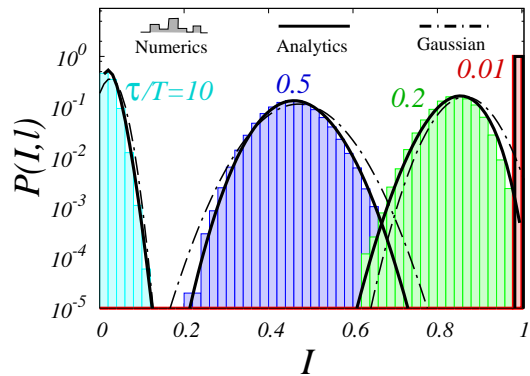}
    \caption{\label{pnfig} Large deviation function. The probability distribution $P(I,l=50)$ of the current $I$ for different values of the correlation time $\tau$ or, equivalently, $\varepsilon=e^{-T/2\tau}$. The full lines show the saddle-point approximation (\ref{pnformula}), while dashed lines correspond to a Gauss distribution with mean and variance (the second cumulant) obtained from Eq.\ (\ref{eq:cumulants}). The histograms were assembled from 50.000 numerical realizations.}
  \end{center}
\end{figure}

\textit{Large deviation function}.--- The above statements can be further corroborated by calculating the full distribution function $P(I,l)\rightarrow \int_{-\pi}^{\pi}\frac{d\chi}{2\pi}e^{l\{\ln[\lambda(\chi)]-i\chi I\}}$ of the current $I\equiv n/l$ measured after many periods using a saddle-point approximation \cite{Bar03,Pis04}. This procedure yields
\begin{equation}\label{pnformula}
  P(I,l)\simeq \frac{\left[\frac{\varepsilon(1+I)}{1-I}\right]^l\,\left[\frac{(1-I^2)(1-\varepsilon)^2}{4\varepsilon I^2}\right]^{Il}}{\sqrt{\pi I(1-I^2)l}}.
\end{equation}
In Fig.\ \ref{pnfig} we show our analytic result together with numerical simulations. For comparison, we also show a Gauss distribution with mean and variance (second cumulant) obtained from Eq.\ (\ref{eq:cumulants}). The analytic result describes our numerics very well. Close to the phase noise regime ($\varepsilon\simeq 0$), the distribution function is clearly left-skewed in accordance with the third cumulant being negative in Fig.\ \ref{figcumulant}. Additionally, the distribution is super-gaussian with heavy tails in comparison to the Gauss distribution. As the shot noise regime is approached, the distribution becomes sub-gaussian and light-tailed. These qualitative changes are reflected in the sign-change of the fourth cumulant seen in Fig.\ \ref{figcumulant}. In the phase noise regime, the distribution is 1 for $I=1$ and zero otherwise. This can be understood by considering the expected mean number of ``cycle missing'' events in our numerical simulations. After $l=50$ periods of the driving, the expected mean number $2le^{-T/2\tau}$ is still vanishingly small for $\tau/T=0.01$, and we do not observe a single ``cycle missing'' event during 50.000 numerical realizations as seen in Fig.\ \ref{pnfig}. The large deviation function was recently measured in single electron transport through a Coulomb blockade quantum dot \cite{Fri10}, and the results presented in Fig.\ \ref{pnfig} may serve as an experimental test of our model in similar measurements on a quantum capacitor.

\textit{Conclusions}.--- We have analyzed the quantum capacitor and found analytically the noise spectrum which fully accounts for recent measurements. We have characterized the accuracy of the capacitor as a single electron source through calculations of the counting statistics and found that the failure rate can be  arbitrarily small under favorable operating conditions. Our results are important for possible applications of quantum capacitors in nano-scale electronics operating at giga-hertz frequencies and our predictions may be tested in future experiments.

\textit{Acknowledgements}.--- We thank G.\ F\`eve, M.\ Moskalets and S.\ E.\ Nigg for useful discussions. The work was supported by the Swiss NSF, MaNEP, and the Carlsberg Foundation.

\end{document}